# High-Throughput Screening of Transition Metal Binuclear Site for $N_2$ Fixation

Xingshuai Lv,[a,b] Wei Wei,[*,a] Baibiao Huang,[a] Ying Dai[*,a] and Thomas Frauenheim[*,bcd]

[a] School of Physics, State Key Laboratory of Crystal Materials, Shandong University, 250100 Jinan, China

[b] Bremen Center for Computational Materials Science, University of Bremen, 2835 Bremen, Germany

[c] Beijing Computational Science Research Center (CSRC), 100193 Beijing, China

[d] Shenzhen JL Computational Science and Applied Research Institute, 518110 Shenzhen, China

[*]Corresponding authors:

weiw@sdu.edu.cn (W. Wei)

daiy60@sdu.edu.cn (Y. Dai)

thomas.frauenheim@bccms.uni-bremen.de (T. Frauenheim)

**ABSTRACT:** Great enthusiasm in single-atom catalysts (SACs) for the $N_2$ reduction reaction (NRR) has been aroused by the discovery of Metal (M)−$N_x$ as a promising catalytic center. However, the performance of available SACs, including poor activity and selectivity, is far away from the industrial requirement because of the inappropriate adsorption behaviors of the catalytic centers. Through the first-principles high-throughput screening, we find that the rational construction of Fe−Fe dual-atom centered site distributed on graphite carbon nitride ($Fe_2$/$g$-CN) compromises the ability to adsorb $N_2H$ and $NH_2$, achieving the best NRR performance among 23 different transition metal (TM) centers. Our results show that Fe$_2$/$g$-CN can achieve a Faradic efficiency of 100% for NH$_3$ production. Impressively, the limiting-potential of Fe$_2$/$g$-CN is estimated as low as −0.13 V, which is hitherto the lowest value among the reported theoretical results. Multiple-level descriptors (excess electrons on the adsorbed N$_2$ and integrated-crystal orbital Hamilton population) shed light on the origin of NRR activity from the view of energy, electronic structure, and basic characteristics. As a ubiquitous issue during electrocatalytic NRR, ammonia contamination originating from the substrate decomposition can be surmounted. Our predictions offer a new platform for electrocatalytic synthesis of NH$_3$, contributing to further elucidate the structure−performance correlations.

1. INTRODUCTION

Large-scale ammonia synthesis is indispensable for the mass production of fertilizers and chemicals.[1] Traditionally, the predominant method for industrial-scale ammonia synthesis has been the Haber−Bosch process, which carries many disadvantages including emissions of considerable CO$_2$ and energy-intensive consumption.[2,3] As such, electrochemical nitrogen reduction reaction (NRR) at ambient conditions has spurred a growing interest, because it is more cost-effective, efficient, sustainable and environmentally friendly.[4−10] A simple description of the electrode reaction is that N$_2$ molecules are adsorbed on electrode surfaces and then reduced by electrons supplied from an external circuit with simultaneous proton addition to form ammonia. Major challenges in electrochemical ammonia synthesis are related to the low activity and selectivity of currently available electrocatalysts (Most of the reported NRR electrocatalysts are transition metal (TM)-based materials, thanks to their advantageous coupling of empty and occupied d orbitals) for $N_2$ reduction.[11–15] The rational design of active, efficient and durable NRR electrocatalysts is still stumbling block. In this scenario, boosting the electrocatalytic NRR efficiencies remains one of the greatest challenges in this field.[16]

Single atom catalysts (SACs), especially single transition−metal atom materials featured by TM−N coordinate sites that mimic biological porphyrin catalysts, have emerged as a brand-new class of heterogeneous catalysts and have attracted considerable interest due to their maximized atomic utilization of catalytically active metals.[17–24] Beyond SACs, diatom catalysts (DACs), in which metal dimers are uniformly dispersed on substrate with elevating catalytic properties, have recently emerged as an extended family member. On account of the synergetic interaction, dimer sites hold special advantages over single-atom sites in term of reducing energy input along alternative reaction paths. Compared with their single-atom counterparts, several DACs were predicted to have better catalytic performance for hydrogen and oxygen reduction reaction,[25,26] $CO_2$ reduction reaction,[27] CO oxidation[28] and NRR[29–32] than SACs. By coupling different metal centers with a variety of supports, DACs could contribute to broad prospect for the search of electrocatalyst candidates. These inspired us to check if the "ligand−metal" concept from homogeneous molecular catalysis could be adopted to guide the design and optimization of TM−N SACs for NRR electrocatalysts.

On the basis of this concept, we propose an active center of di-TM pair anchored on a heterogeneous substrate, $g$-CN, to construct $TM_2$/$g$-CN DACs (**Figure 1a**), which can be deemed as N$_2$ capture and conversion catalysts, based on three considerations: (i) *g*-CN featured affluent porosity has been synthesized by a solvothermal method;[33] (ii) the strong coupling between single atoms and the porous characteristics not only stabilizes the catalyst systems, but also triggers electron redistribution to boost the catalytic activity for NRR; (iii) TM$_2$/*g*-CN are composed of two TMN$_2$ units, belonging to the family of M−N materials that are highly active toward the NRR. Our results show that, although a high tendency of TM−adsorbate electron donation can promote the hydrogenation of *N$_2$ to *N$_2$H, it would also severely hamper the *NH$_2$ to *NH$_3$ conversion due to the strong TM−N bonding. This electron-donation concept enables Fe$_2$/*g*-CN the best catalytic performance for NRR with the lowest limiting-potential of −0.13 V ever reported.

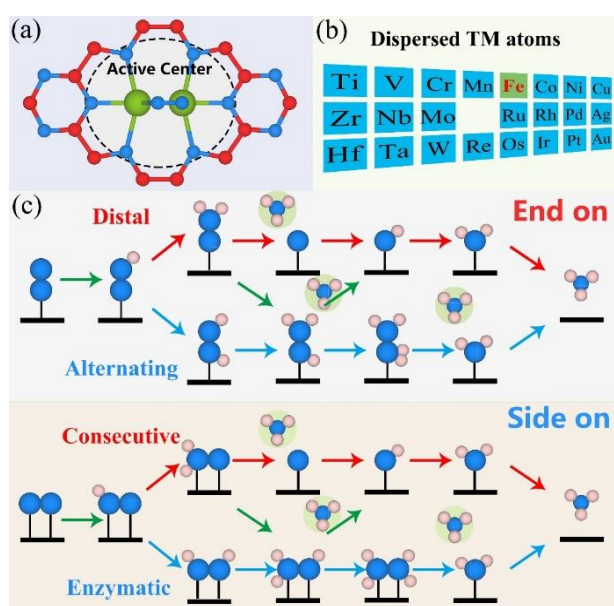

**Figure 1** (a) Crystal structure of N$_2$ coordinated with two heterogeneous TMN$_3$ active sites. (b) 23 kinds of transition metals are taken into account in this work. (c) Schematic depiction of distal, alternating, consecutive, enzymatic and mixed mechanisms for N$_2$ reduction to NH$_3$. Blue and pink atoms represent N and H atoms, respectively.

## 2. COMPUTATIONAL METHODS

All the spin-polarized density functional theory (DFT) calculations were carried out by using Vienna ab initio simulation package[34,35] with projected augmented wave (PAW) method.[36] A cut-off energy of 450 eV was adopted and the Perdew−Burke−Ernzerhof (PBE) exchange-correlation functional[37] was employed. The DFT−D2 by Grimme[38] was used to account for dispersion interactions for structural optimization. All the structures were fully relaxed until the forces on each atom were less than 0.02 eV/Å and the convergence criterion for the electronic structure iteration was set to $10^{-5}$ eV. The *k*-points in the Brillouin zone were sampled with a 5×5×1 grid. To avoid the interaction between periodic images, a vacuum space over 20 Å was used.

Gibbs free energy change ($\Delta G$) for each elementary step was calculated based on the computational hydrogen electrode (CHE) model[39] by the following equation

$$\Delta G = \Delta E + \Delta E_{ZPE} - T\Delta S$$

where $\Delta E$, $\Delta E_{ZPE}$ and $\Delta S$ are the reaction energy from DFT calculation; the changes in zero-point energy and entropy, respectively, which are obtained from the vibrational frequency calculations; *T* is the temperature (*T* = 298.15 K). The computational details for durability, Faradic efficiency and synthesis of $Fe_2$/*g*-CN are given in the **Supporting Information (SI)**.

## 3. RESULTS AND DISCUSSION

**3.1. Screening of the DACs.**

To fully eluciadte the effect of metal ceters on the NRR intrinsic activity, we screened 23 metal centers including all metals in 3d, 4d and 5d blocks as shown in **Figure 1b**. The $N_2$ can be reduced through both end-on and side-on adsorption configurations, including distal, alternating, enzymatic and consecutive pathways as shown in **Figures 1c**. The limiting potential, defined as the lowest negative potential at which the pathway becomed exergonic, was used to evaluate the intrinsic activity of NRR catalysts.[40–47] As the NRR is a complicated process, efficient screening descriptors are therefore desirable. For this purpose, we present a "Five-step" strategy as shown in **Figure 2a**: First, DACs should possess high thermodynamic stability ($\Delta E_b < 0$ eV) and feasibility ($E_f < 0$ eV); Secondly, $N_2$ should be sufficiently activated with the $\Delta G_{*N2} < -0.3$ eV; Thirdly and Fourthly, to guarantee low energy cost, the $\Delta G$ of the first and last hydrogenation step (the most likely limiting steps) should be as low as possible with $\Delta G_{*N2H}$ and $\Delta G_{*NH3} < 0.55$ eV (the best catalyst Ru); Finally, to guarantee the high selectivity of NRR, the maximum $\Delta G$ for NRR should be much lower than that of the competing hydrogen evolution reaction (HER).

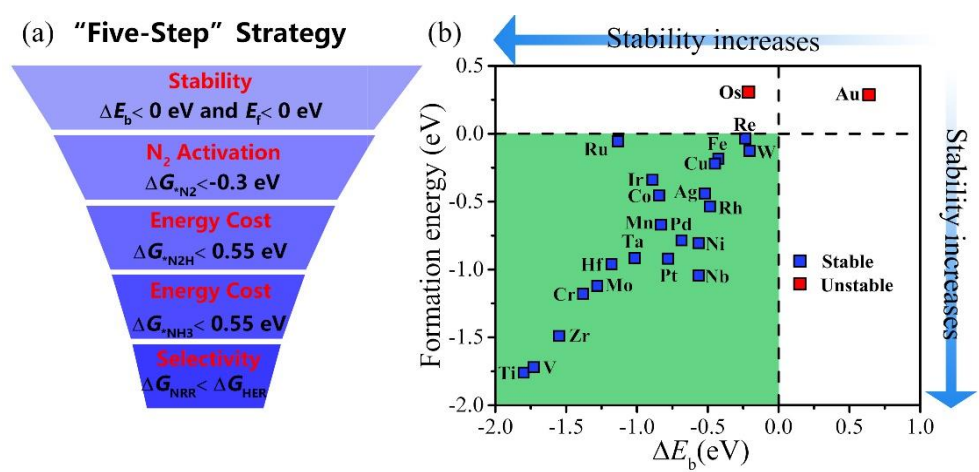

**Figure 2** (a) The proposed "five-step" strategy for screening NRR candidates. (b) Computed $\Delta E_b$ and $E_f$ of metal atoms on *g*-CN.

For DACs, one major issue is that isolated metal atoms tend to aggregate into metal clusters due to the high surface free energy. Therefore, we accessed the stability and feasibility of DACs by the formation energy ($E_f$) and the energy difference ($\Delta E_b$) between binding energy ($E_b$) of TM atoms on *g*-CN and the cohesive energy of TM atoms ($E_{coh}$), see **Table S1**. According to the definition, the negative $\Delta E_b$ indicates that TM atoms on *g*-CN are energetically more favorable than in the bulk form, so the TM atoms are unlikely to aggregate into clusters. In addition, systems with $E_f < 0$ eV are considered as thermodynamically stable and accessible. As shown in **Figure 2b**, most of the DACs can satisfy the criteria, which suggests the reliability and feasibility of our approach.

A key step during NRR is the adsorption and activation of the insert $N_2$ and the $\Delta G_{*N2}$ are shown in **Figure 3a**, which present a dependence on the number of $d$ electrons of metal atoms. For the same row of metal atoms, $\Delta G_{*N2}$ are gradually increase from left to right. This can be understood by the bonding characteristic between $N_2$ and TM atoms. Based on the "electron acceptance and donation" concept, the TM atoms could provide unoccupied $d$ orbitals to accept lone-pair electrons from $N_2$, and meanwhile, the occupied d orbitals of TM atoms could back-donate electrons into the antibonding orbitals of $N_2$. As such, TM−N bonds are strengthened while simultaneously N≡N bonds are weakened. For early TM atoms, their $d$ electrons is more likely to donate to the $N_2$ than the late TM atoms, resulting in a much stronger binding strength with $N_2$. Following criterion 1, Ti, Fe, Co, Zr, Nb, Ru, Hf, Ta, W, Re and Os are singled out. As shown in **Figure 3b**, following the criteria 2, Ti and Fe are further screened out. Finally, taken the selectivity into account, $Fe_2/g$-CN is deemed as the only eligible electrocatalyst candidate for the NRR satisfying all five screening criteria (**Figures 3c** and **d**).

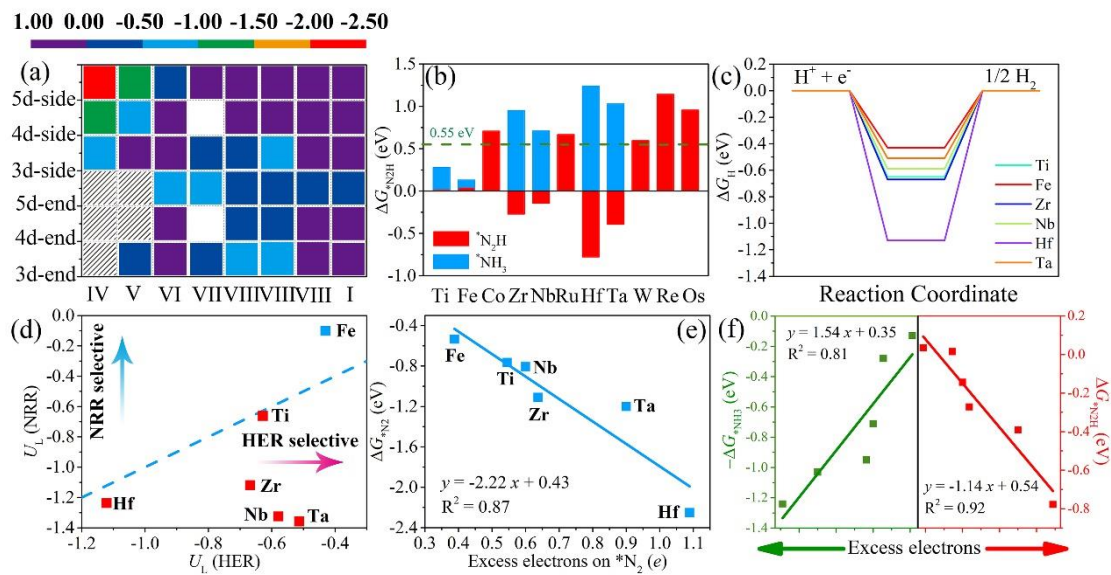

**Figure 3** (a) Adsorption energies of $N_2$ on $TM_2$/$g$-CN via both end-on and side-on patterns. (b) Free energy changes of $^*N_2 + H^+ + e^- \rightarrow {}^*N_2H$ (red) and $^*NH_2 + H^+ + e^- \rightarrow {}^*NH_3$ (blue) on $TM_2$/$g$-CN. (c) Free energy changes for hydrogen evolution reaction. (d) Limiting potentials for NRR ($U_L$(NRR)) and HER ($U_L$(HER)) illustrating the NRR selectivity of $TM_2$/$g$-CN. When $U_L$(NRR) < $U_L$(HER), $TM_2$/$g$-CN are selective for NRR. (e) Free energy changes of $^*N_2$ adsorbed on $TM_2$/$g$-CN as a function of the excess electrons on $^*N_2$, where the shadows represent that $N_2$ cannot be adsorbed on $TM_2$/$g$-CN via this pattern. (f) Active volcano curve of $\Delta G_{*N2H}$ (red) and $\Delta G_{*NH3}$ (green) as a function of excess electrons on $^*N_2$.

In order to further reveal the reaction mechanism, we established the activity trends of various DACs as a function of some parameters. **Figure 3e** elucidates that the degree of electron donation from TM sites to $N_2$ for facilitating the activation of $N_2$, quantified by the excess electron on $N_2$, is linearly correlated with the binding strength of TM−$N_2$. The late TM atoms belonging to the same row are reluctant to donate electrons and thus displays poorer ability to activate $N_2$. **Figure 3f** shows the volcano plot of $\Delta G_{*N2H}$ and $\Delta G_{*NH3}$ as a functional of excess electron on $N_2$. Clearly, DACs that binding $N_2$ too weakly are limited by the first hydrogenation step, whereas those with too strong binding strength (poisoning effect due to the strong interaction between the DACs and $^*N_2H$) are limited by the protonation of $^*NH_2$ to form $^*NH_3$. In addition, the occurrence of large spin-polarization in Fe could force $^*N_2$ to be of radical nature (with unpaired electron) that is active for hydrogenation (**Figures S1a** and **b**).[48,49] In this scenario, Fe$_2$/*g*-CN is therefore validated as the most promising electrocatalyst for the NRR among all candidates.

To gain deep insights into the d−π$^*$ interaction, we performed the partial density of states (PDOS) and the integrated-crystal orbital Hamilton population (ICOHP) analysis. Here, the ICOHPs (which is obtained by calculating the energy integral up to the highest occupied energy level) could be a quantitative indicator of N$_2$ activation degree, and a more negative value of ICOHP implies a less activated N$_2$ molecule. As shown in **Figure 4**, the strong d−π$^*$ coupling can activate the adsorbed N$_2$ to be radical-like, which is ready for hydrogenation. **Figures 4b-g** show that when d orbital is less filled (Ti, Zr and Hf), the filling of antibonding orbital population of N$_2$ decreases, which is in accordance with the increase of binding strength and indicates that N≡N bonding is weaker. The weakened N≡N bond is also verified by a less negative ICOHP with a less antibonding orbital filling. More importantly, we discovered that there is a good linear correlation between ICOHP and excess electrons on adsorbed N$_2$ (**Figure 4h**). This linear correlation gives a quantitative explanation for the role of different metal centers in determining the bonding/antibonding orbital populations, which is the origin of the best catalytic performance of Fe$_2$/*g*-CN.

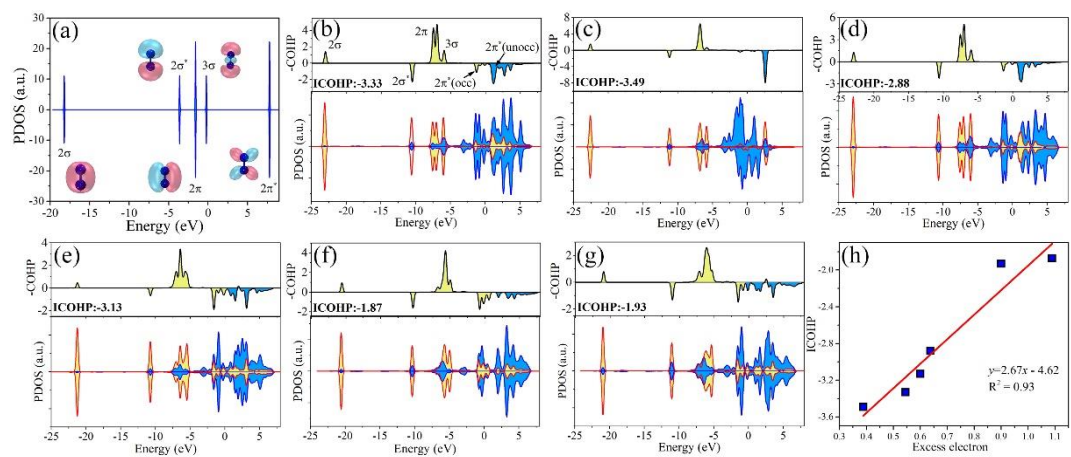

**Figure 4** (a-g) Molecular orbitals of free N$_2$, and the computed partial density of states (PDOS) and the crystal orbital Hamilton populations (COHPs) of N$_2$ on TM$_2$/*g*-CN (TM = Ti, Fe, Zr, Nb, Hf, and Ta). The bonding and antibonding states in COHP are depicted by yellow and cyan, respectively. (h) Illustration of the correlation between integrated COHP (ICOHP) and the excess electrons on *$N_2$.

## 3.2. The NRR performance of Fe$_2$/g-CN.

Since the catalytic activity of NRR is governed by multiple reaction intermediates, the detailed reaction pathways and activity trends will be systematically evaluated. All of the elementary steps associated with distal and alternating pathways in **Figure 1c** are taken into consideration. Thermodynamic results show that NRR processes prefer to adopt the mixed pathway with the following intermediates: *$N_2$ → *NNH → *$NNH_2$ → *$NHNH_2$ → *$NHNH_3$ → *$NH_2$ → *$NH_3$. The last hydrogenation step is identified as the potential limiting step (PDS), which is unlike most of the TM-based catalysts that the PDS is imposed by the first hydrogenation step.[13] Importantly, **Figures 5a** and **5b** summarize the free energy change and adsorption configurations for various intermediates on Fe$_2$/g-CN, indicating that the limiting potential is as low as −0.13 V. It is worth noting that −0.13 is well below that on the Ru(0001) surface (1.08 V) and hitherto the lowest value among the reported theoretical results,[10] indicative of the lowest energy cost for NRR electrocatalysts ever reported.

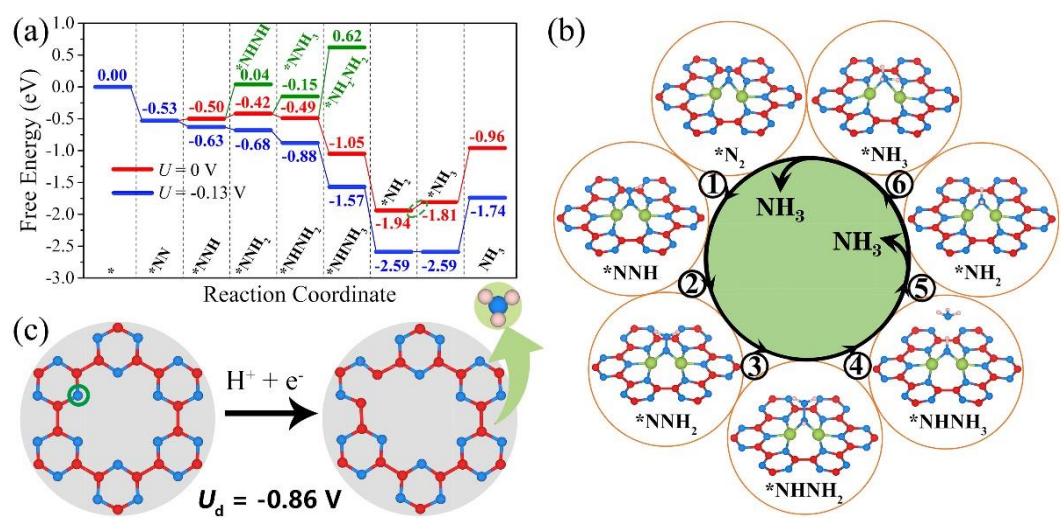

**Figure 5** (a) Free energy diagrams and (b) corresponding adsorption structures of reaction intermediates for N$_2$ reduction on Fe$_2$/g-CN through the minimum energy pathway. (c) Decomposition energy of substrate g-CN.

Another issue that limits NRR performance in acidic condition is the Faradic efficiency (FE), closely related to the competitive HER. According to the equation of FE in the SI, the FE of $Fe_2$/$g$-CN can reach up to 100%, which is much better than those for pure transition metal surfaces.[13] Therefore, by combining the activity (limiting potential) and selectivity (FE) analysis, $Fe_2$/$g$-CN holds the potential for NRR and could be further examined experimentally.

**3.3. Practical application of $Fe_2$/$g$-CN as electrocatalyst for NRR.**

For the practical implementations of electrocatalysts, $Fe_2$/$g$-CN need to have good electric conductivity, durability and high synthetic accessibility. As shown in **Figure S2a**, pristine $g$-CN is a semiconductor with a large band gap of 3.21 eV, which is consistent with previous result.[40,50] After depositing Fe atoms, $Fe_2$/$g$-CN presents metallic properties due to the electron redistribution between the Fe atom partially occupied $d$ bands and $g$-CN (**Figure S2b**), endowing $Fe_2$/$g$-CN excellent electric conductivity to expedite charge transport and improve electrocatalytic NRR activity. In addition, the stability and durability of $Fe_2$/$g$-CN are evaluated based on ab initio molecular dynamics (AIMD) simulations and thermodynamics. In accordance with the results shown in **Figure S3**, Fe atoms are still firmly anchored in the hole at 500 K after 10 ps. Because nitrogen atoms are contained in $g$-CN, these nitrogen atoms may cause ammonia contamination as a result of decomposition of $g$-CN substrate, which is a ubiquitous issue during electrocatalytic NRR. Therefore, we examined the durability of $g$-CN substrate as presented in **Figure 5c** (detailed computational methods are given in the SI). According to previous studies, the NRR electrocatalysts should have more positive NRR limiting potential to avoid risks of contamination due to the decomposition of substrate.[17] Here, the results show that decomposition potential for $g$-CN substrate is −0.86 V, which is much lower than that of the NRR (−0.13 V). Therefore, Fe$_2$/*g*-CN can be validated as stable and durable electrocatalysts.

For the synthesis of DACs, the wet chemistry method is a suitable way to achieve highly dispersed single atoms.[51,52] Inspired by previous studies,[53,54] using FeCl$_2$ as the metal precursor, we examined the feasibility for the experimental realization of Fe$_2$/*g*-CN by computing the energy profile of the synthetic route as shown in **Figures S4** and **S5**. All the reaction steps can easily occur because they are either spontaneous or only slightly endothermic (**Figure S5**). Thus, the synthesis of Fe$_2$/*g*-CN through such an synthetic route is likely realized experimentally.

## 4. CONCLUSIONS

In summary, we conducted extensive DFT calculations to evaluate the potential of TM-DACs supported on *g*-CN in electrochemical NRR and revealed the reaction mechanisms for di-TM center catalysis. Upon a "five-step" screening strategy, Fe$_2$/*g*-CN is singled out as the most promising candidate with low energy cost and selectivity, on which the limiting potential for NRR is as low as −0.13 V. The last hydrogenation step is identified as the PDS, depending on the degree of TM-to-N$_2$ electron donation. Specifically, a higher tendency of electron donation can enable efficient N$_2$ activation, whereas a low tendency contributes to the formation of *NH$_3$, thereby reducing the energy input. In this scenario, a certain moderate degree of electron donation, tunable by different TM and ligands, would be desirable for NRR. Besides, the examination of the selectivity (FE of 100%), stability and durability as well as the synthetic accessibility support the great potential of Fe$_2$/*g*-CN as electrocatalysts for NRR. This systematic work offer new insights toward the discovery of more efficient TM-DACs for NRR.

■ ACKNOWLEDGMENTS

This work is supported by the National Natural Science foundation of China (No. 11374190 and 21333006), the Taishan Scholar Program of Shandong Province and the Young Scholars Program of Shandong University (YSPSDU). TF acknowledges funding from DFG-RTG 2247.